\documentclass[codesnippet]{jss}
\usepackage[latin1]{inputenc}
\usepackage{amsmath}
\usepackage{amsfonts}
\usepackage{amssymb}
\usepackage{graphicx}
\usepackage{color}
\usepackage{amsthm}

\def\log#1{\text{log}\left(#1\right)}

\newcommand{\bbeta}{\boldsymbol\beta}

\newcommand{\xx}{\mathbf{x}}

\newcommand{\XX}{\mathbf{X}}

\newcommand{\g}{\mathbf{g}}
\newcommand{\G}{\mathbf{G}}

\newcommand{\HH}{\mathbf{H}}
\newcommand{\hh}{\mathbf{h}}

\newcommand{\pp}{\mathbf{p}}
\newcommand{\qq}{\mathbf{q}}
\newcommand{\PP}{\mathrm{P}}

\author{Alireza S. Mahani\\ Scientific Computing Group \\ Sentrana Inc. \And
Mansour T.A. Sharabiani\\ National Heart and Lung Institute \\ Imperial College London}
\title{Expander Framework for Generating High-Dimensional GLM Gradient and Hessian from Low-Dimensional Base Distributions: \proglang{R} Package \pkg{RegressionFactory}}

\Plainauthor{Alireza S. Mahani, Mansour T.A. Sharabiani} 
\Plaintitle{Expander Functions for Generating High-Dimensional Gradient and Hessian from Low-Dimensional Base Distributions: R Package RegressionFactory} 
\Shorttitle{Regression Expander Functions: \proglang{R} Package \pkg{RegressionFactory}} 

\Abstract{ 
The \proglang{R} package \pkg{RegressionFactory} provides expander functions for constructing the high-dimensional gradient vector and Hessian matrix of the log-likelihood function for generalized linear models (GLMs), from the lower-dimensional base-distribution derivatives. The software follows a modular implementation using the chain rule of derivatives. Such modularity offers a clear separation of case-specific components (base distribution functional form and link functions) from common steps (e.g., matrix algebra operations needed for expansion) in calculating log-likelihood derivatives. In doing so, \pkg{RegressionFactory} offers several advantages: 1) It provides a fast and convenient method for constructing log-likelihood and its derivatives by requiring only the low-dimensional, base-distribution derivatives, 2) The accompanying definiteness-invariance theorem allows researchers to reason about the negative-definiteness of the log-likelihood Hessian in the much lower-dimensional space of the base distributions, 3) The factorized, abstract view of regression suggests opportunities to generate novel regression models, and 4) Computational techniques for performance optimization can be developed generically in the abstract framework and be readily applicable across all the specific regression instances. We expect \pkg{RegressionFactory} to facilitate research and development on optimization and sampling techniques for GLM log-likelihoods as well as construction of composite models from GLM lego blocks, such as Hierarchical Bayesian models.
}
\Plainkeywords{negative definiteness, regression, optimization, sampling, monte carlo markov chain, hierarchical bayesian} 

\Address{
Alireza S. Mahani\\
Scientific Computing Group\\
Sentrana Inc.\\
1725 I St NW\\
Washington, DC 20006\\
E-mail: \email{alireza.mahani@sentrana.com}\\
}

\usepackage{Sweave}
\begin{document}
\Sconcordance{concordance:RegressionFactory.tex:RegressionFactory.Rnw:%
1 95 1 1 0 4 1 1 4 127 1 1 14 16 0 1 2 8 1 1 40 42 0 1 2 52 1 %
1 2 4 0 1 2 1 4 6 0 1 2 1 9 11 0 1 2 1 6 8 0 1 2 2 1 1 2 1 0 5 %
1 3 0 2 2 1 0 1 2 1 0 4 1 1 5 4 0 2 1 11 0 2 2 1 0 5 1 1 3 2 0 %
1 4 6 0 2 2 1 0 1 4 6 0 2 2 1 0 1 1 1 8 7 0 1 1 3 0 2 2 12 0 1 %
1 12 0 1 2 1 4 3 0 1 1 3 0 1 2 2 1 1 -4 1 8 8 1 1 2 4 0 1 2 1 %
7 9 0 2 2 1 0 7 1 3 0 1 2 1 3 2 0 2 1 1 2 6 0 2 1 3 0 1 2 1 3 %
2 0 1 1 1 5 4 0 3 1 10 0 1 1 6 0 1 2 1 0 1 1 1 5 4 0 3 1 10 0 %
1 1 11 0 1 2 15 1 1 2 1 0 4 1 3 0 1 2 1 4 3 0 1 1 1 4 3 0 1 1 %
11 0 1 2 17 1}



\section{Introduction}\label{section-intro}
Generalized Linear Models (GLMs)~\citep{mccullagh1989generalized} are one of the most widely-used classes of models in statistical analysis, and their properties have been thoroughly studied and documented (see, for example, \cite{dunteman2006introduction}). Model training and prediction for GLMs often involves Maximum-Likelihood estimation (frequentist approaches) or posterior density estimation (Bayesian approaches), both of which require application of optimization or MCMC sampling techniques to the log-likelihood function or some function containing it. Differentiable functions often benefit from optimization/sampling algorithms that utilize the first and/or second derivative of the function~\citep{press2007numerical}. With proper choice of link functions, many GLMs have log-likelihood functions that are not only twice-differentiable, but also globally-concave~\citep{gilks1992adaptive}, making them ideal candidates for optimization/sampling routines that take advantage of these properties. For example, the most common optimization approach for GLMs is Iterative Reweighted Least Squares (IRLS)~\citep[Section~6.8.1]{gentle2007matrix}. IRLS is a disguised form of Newton-Raphson optimization~\citep{wright1999numerical}, which uses both the gradient and Hessian of the function, and relies on global concavity for convergence. When Hessian is too expensive to calculate or lacks definiteness, other optimization techniques such as conjugate gradient~\citep[Section~10.6]{press2007numerical} can be used, which still require the first derivative of the function. Among MCMC sampling algorithms, Adaptive Rejection Sampler~\citep{gilks1992adaptive} uses the first derivative and requires concavity of the log-density. Stochastic Newton Sampler~\citep{qi2002hessian,mahani2014sns}, a Metropolis-Hastings sampler using a locally-fitted multivariate Gaussian, uses both first and second derivatives and also requires log-concavity. Other techniques such as Hamiltonian Monte Carlo (HMC)~\citep{neal2011mcmc} use the first derivative of log-density, while their recent adaptations can use second and even third derivative information to adjust the mass matrix to local space geometry~\citep{girolami2011riemann}. Efficient implementation and analysis of GLM derivatives and their properties, therefore, is a key component to our ability to build probabilistic models using the powerful GLM framework.

The \proglang{R} package \pkg{RegressionFactory} contributes to computational research and development on GLM-based statistical models by providing an abstract framework for constructing, and reasoning about, GLM-like log-likelihood functions and their derivatives. Its modular implementation can be viewed as code factorization using the chain rule of derivatives~\citep{apostol1974mathematical}. It offers a clear separation of generic steps (expander functions) from model-specific steps (base functions). New regression models can be readily implemented by supplying their base function implementation. Since base functions are in the much lower-dimensional space of the underlying probability distribution (often a member of the exponential family with one or two parameters), implementation of their derivatives is much easier than doing so in the high-dimensional space of regression coefficients. A by-product of this code refactoring using the chain rule is an invariance theorem governing the negative definiteness of the log-likelihood Hessian. The theorem allows this property to be studied in the base-distribution space, again a much easier task than doing so in the high-dimensional coefficient space. The modular organization of \pkg{RegressionFactory} also allows for performance optimization techniques to be made available across a broad set of regression models. This is particularly true for optimizations applied to expander functions, but also applies to base functions since they share many concepts and operations across models. \pkg{RegressionFactory} contains a lower-level set of tools compared to the facilities provided by mainstream regression utilities such as the \code{glm} command in \proglang{R}, or the package \pkg{dglm}~\citep{dunn2014dglm} for building double (varying-dispersion) GLM models. Therefore, in addition to supporting research on optimization/sampling algorithms for GLMs as well as research on performance optimization for GLM derivative-calculation routines, exposing the log-likelihood derivatives using the modular framework of \pkg{RegressionFactory} allows modelers to construct composite models from GLM lego blocks, including Hierarchical Bayesian models~\citep{gelman2006data}.

The rest of the paper is organized as follows. In Section~\ref{section-theory}, we begin with an overview of GLM models and arrive at our abstract, and expanded, representation of GLM log-likelihoods (\ref{subsection-overview}). We then apply the chain rule of derviatives to this abstract expression to derive two equivalent sets of factorized equations (compact and explicit forms) for computing log-likelihood gradient and Hessian using their base-function counterparts (\ref{subsection-chain-rule}). We use the explicit forms of the equations to prove a negative-definiteness invariance theorem for the log-likelihood Hessian (\ref{subsection-invariance}). Section~\ref{section-implement} discusses the implementation of the aforementioned factorized code in \pkg{RegressionFactory} using the expander functions (\ref{subsection-expanders}) and the base functions (\ref{subsection-base-dist}). In Section~\ref{section-using}, we illustrate the use of \pkg{RegressionFactory} using examples from single-parameter and multi-parameter base functions. Finally, Section~\ref{section-summary} contains a summary and discussion.

\section{Theory}\label{section-theory}
In this section we develop the theoretical foundation for \pkg{RegressionFactory}, beginning with an overview of GLM models.
\subsection{Overview of GLMs}\label{subsection-overview}
In GLMs, response variable\footnote{To simplify notation, we assume that response variable is scalar, but generalization to vector response variables is straightforward.} $y$ is assumed to be generated from an exponential-family distribution, and its expected value is related to linear predictor $\xx^t \bbeta$ via the link function $g$:
\begin{equation}\label{equation-glm}
g(\mathrm{E}(y)) = \xx^t \bbeta.
\end{equation}
where $\xx$ is the vector of covariates and $\bbeta$ is the vector of coefficients. For single-parameter distributions, there is often a simple relationship between the distribution parameter and its mean. Combined with Equation~\ref{equation-glm}, this is sufficient to define the distribution in terms of the linear predictor, $\xx^t \bbeta$. For many double-parameter distributions, the distribution can be expressed as
\begin{equation}\label{equation-dglm}
f_Y(y; \theta, \Phi) = \exp \{ \frac{y \theta - B(\theta)}{\Phi} + C(y,\Phi) \}
\end{equation}
where range of $y$ does not depend on $\theta$ or $\Phi$. This function can be maximized with respect to $\theta$ without knowledge of $\Phi$. Same is true if we have multiple conditionally-independent data points, where log-likelihood takes a summative form. Once $\theta$ is found, we can find $\Phi$ (dispersion parameter) through maximization or method of moments, as done by \code{glm} in \proglang{R}. Generalization to varying-dispersion models is offered in the \proglang{R} package \pkg{dglm}, where both mean and dispersion are assumed to be linear functions of covariates. In \pkg{dglm} estimation is done iteratively by alternating between an ordinary GLM and a dual GLM in which the deviance components of the ordinary GLM appear as responses~\citep{smyth1989generalized}.

In \pkg{RegressionFactory}, we take a more general approach to GLMs that encompasses the \code{glm} and \code{dglm} approaches but is more flexible. Our basic assumption is that log-density for each data point can be written as:

\begin{equation}
\log {\PP (y \,|\, \{\ \xx^j \}_{j=1,\dots,J})} = f(<\xx^1,\bbeta^1>, \dots, <\xx^J,\bbeta^J> , y)
\end{equation}
where $<a,b>$ means inner product of vectors $a$ and $b$. Note that we have absorbed the nonlinearities introduced through one or more link functions into the definition of $f$. For $N$ conditionally-independent observations $y_1,\dots,y_N$, the log-likelihood as a function of coefficients $\bbeta^j$ is given by:
\begin{equation}\label{eq-loglike}
L(\bbeta^1, \dots, \bbeta^J) = \sum_{n=1}^N f_n(<\xx_n^1, \bbeta^1>, \dots, <\xx_n^J, \bbeta^J>),
\end{equation}
where we have absorbed the dependence of each term on $y_n$ into the indexes of the base functions $f_n(u^1,\dots,u^J)$. With proper choice of nonlinear transformations, we can assume that the domain of $L$ is $\mathbb{R}^{\sum_j K^j}$, where $K^j$ is the dimensionality of $\bbeta^j$.

This view of GLMs naturally unites single-parameter GLMs such as Binomial (with fixed number of trials) and Poisson, constant-dispersion two-parameter GLMs (e.g. normal and Gamma), varying-dispersion two-parameter GLMs (e.g. heteroscedastic normal regression), and multi-parameter models such as multinomial logit. It can motivate new GLM models such as geometric (see Section~\ref{subsection-geometric}) and exponential, and can even include survival models (see, e.g., \pkg{BSGW} package~\citep{mahani2014bsgw}). Several examples are discussed in Section~\ref{section-using}.

Our next step is to apply the chain rule of derivatives to Equation~\ref{eq-loglike} to express the high-dimensional ($\sum_j K^j$) derivatives of $L$ in terms of the low-dimensional ($J$) derivatives of $f_n$'s. We will see that the resulting expressions offer a natural way for modular implementation of GLM derivatives.

\subsection{Application of chain rule}\label{subsection-chain-rule}
First, we define our notation for representing derivative objects. We concatenate all $J$ coefficient vectors, $\bbeta^j$'s, into a single $\sum_j K^j$-dimensional vector, $\bbeta$:
\begin{equation}
\bbeta \equiv (\bbeta^{1,t}, \dots, \bbeta^{J,t})^t.
\end{equation}
The first derivative of log-likelihood can be written as:
\begin{equation}
\G(\bbeta) \equiv \frac{\partial L}{\partial \bbeta} = ((\frac{\partial L}{\partial \bbeta^1})^t, \dots, (\frac{\partial L}{\partial \bbeta^J})^t)^t,
\end{equation}
where
\begin{equation}
(\frac{\partial L}{\partial \bbeta^j})^t \equiv (\frac{\partial L}{\partial \beta_1^j}, \dots, \frac{\partial L}{\partial \beta_{K^j}^j}).
\end{equation}
For second derivatives we have:
\begin{equation}
\HH(\bbeta) \equiv \frac{\partial^2 L}{\partial \bbeta^2} = \left[ \frac{\partial^2 L}{\partial \bbeta^j \partial \bbeta^{j'}} \right]_{j,j'=1,\dots,J},
\end{equation}
where we have defined $\HH(\bbeta)$ in terms of $J^2$ matrix blocks:
\begin{equation}
\frac{\partial^2 L}{\partial \bbeta^j \partial \bbeta^{j'}} \equiv \left[ \frac{\partial L}{\partial \beta_k^j \partial \beta_{k'}^{j'}} \right]_{j=1,\dots,K^j; j'=1,\dots,K^{j'}}
\end{equation}
Applying the chain rule to the log-likelihood function of Equation~\ref{eq-loglike}, we derive expressions for its first and second derivatives as a function of the derivatives of the base functions $f_1,\dots,f_N$:
\begin{equation}\label{eq-gradient}
\frac{\partial L}{\partial \bbeta^j} = \sum_{n=1}^N \frac{\partial f_n}{\partial \bbeta^j} = \sum_{n=1}^N \frac{\partial f_n}{\partial u^j} \xx_n^j = \XX^{j,t} \g^j,
\end{equation}
with
\begin{equation}
\g^j \equiv (\frac{\partial f_1}{\partial u^j}, \dots, \frac{\partial f_N}{\partial u^j})^t,
\end{equation}
and
\begin{equation}
\XX^j \equiv (\xx_1^j, \dots, \xx_N^j)^t.
\end{equation}
Similarly, for the second derivative we have:
\begin{equation}\label{eq-hessian}
\frac{\partial^2 L}{\partial \bbeta^j \partial \bbeta^{j'}} = \sum_{n=1}^N \frac{\partial^2 f_n}{\partial \bbeta^j \partial \bbeta^{j'}} = \sum_{n=1}^N \frac{\partial^2 f_n}{\partial u^j \partial u^{j'}} \, (\xx_n^j \otimes \xx_n^{j'}) = \XX^{j,t} \hh^{jj'} \XX^{j'},
\end{equation}
where $\hh^{jj'}$ is a diagonal matrix of size $N$ with $n$'th diagonal element defined as:
\begin{equation}
h_n^{jj'} \equiv \frac{\partial^2 f_n}{\partial u^j \partial u^{j'}}
\end{equation}
We refer to the matrix form of the Equations~\ref{eq-gradient} and \ref{eq-hessian} as `compact' forms, and the explicit-sum forms as `explicit' forms. The expander functions in \pkg{RegressionFactory} use the compact form to implement the high-dimensional gradient and Hessian (see Section~\ref{subsection-expanders}), while the definiteness-invariance theorem below utilizes the explicit-sum form of Equation~\ref{eq-hessian}.

\subsection{Definiteness invariance of Hessian}\label{subsection-invariance}
\newtheorem{theorem:concavity_1}{Theorem}
\begin{theorem:concavity_1} \label{theorem:concav_1}
If all $f_n$'s in Equation~\ref{eq-loglike} have negative definite Hessians AND if at least one of $J$ matrices $\XX^j \equiv (\xx^j_1, \dots, \xx^j_N)^t$ is full rank, then $L(\bbeta^1,\dots,\bbeta^J)$ also has a negative-definite Hessian.
\end{theorem:concavity_1}
\begin{proof}
To prove negative-definiteness of $\HH (\bbeta)$ (hereafter referred to as $\HH$ for brevity), we seek to prove that $ \pp^t \HH \pp$ is negative for all non-zero $\pp$ in $\mathbb{R}^{\sum_j K^j}$. We begin by decomposing $\pp$ into $J$ subvectors of length $K^j$ each:
\begin{equation} \label{eq:pp}
\pp = (\pp^{1,t}, \dots, \pp^{J,t})^t.
\end{equation}
We now have:
\begin{eqnarray}
\pp^t \HH \pp &=& \sum_{j,j'=1}^J \pp^{j,t} \, \frac{\partial^2 L}{\partial \bbeta^j \partial \bbeta^{j'}} \, \pp^{j'} \\
&=& \sum_{j,j'} \pp^{j,t} \left( \sum_n \frac{\partial^2 f_n}{\partial u^j \partial u^{j'}} \: . \: (\xx_n^j \otimes \xx_n^{j'}) \right) \pp^{j'} \\
&=& \sum_n \sum_{j,j'} \frac{\partial^2 f_n}{\partial u^j \partial u^{j'}} \: \pp^{j,t} \: (\xx_n^j \otimes \xx_n^{j'}) \: \pp^{j'}
\end{eqnarray}
If we define a set of new vectors $\qq_n$ as:
\begin{eqnarray}
\qq_n \equiv \begin{bmatrix} \pp^{1,t} \xx_n^1 & \cdots & \pp^{J,t} \xx_n^J \end{bmatrix},
\end{eqnarray}
and use $\hh_n$ to denote the $J$-by-$J$ Hessian of $f_n$:
\begin{equation}
\hh_n \equiv [ h_n^{jj'} ]_{j,j'=1,\dots,J},
\end{equation}
we can write:
\begin{equation}
\pp^t \HH \pp = \sum_n \qq_n^t \, \hh_n \, \qq_n.
\end{equation}
Since all $\hh_n$'s are assumed to be negative definite, all $\qq_n^t \, \hh_n \, \qq_n$ terms must be non-positive. Therefore, $\pp^t \HH \pp$ can be non-negative only if all its terms are zero, which is possible only if all $\qq_n$'s are zero vectors. This, in turn, means we must have $\pp^{j,t} \xx_n^j = 0,\:\: \forall \, n,j$. In other words, we must have $\XX^j \pp^j = \emptyset,\:\: \forall \, j$. This means that all $\XX^j$'s have non-singleton nullspaces and therefore cannot be full-rank, which contradicts our assumption. Therefore, $\pp^T \HH \pp$ must be negative. This proves that $\HH$ is negative definite.
\end{proof}
Proving negative-definiteness in the low-dimensional space of base functions is often much easier. For single-parameter distributions, we simply have to prove that the second derivative is negative. For two-parameter distributions, and according to Silvester's criterion~\citep{gilbert1991positive}, it is sufficient to show that both diagonal elements of the base-distribution Hessian as well as its determinant are negative. Note that negative-definiteness depends not only on the distribution but also on the choice of link function(s). For twice-differentiable functions, negative-definiteness of Hessian and log-concavity are equivalent~\citep{boyd2009convex}. \cite{gilks1992adaptive} have a list of log-concave distributions and link functions.

\section{Implementation}\label{section-implement}
\begin{figure}
\centering
\includegraphics[scale=1.5]{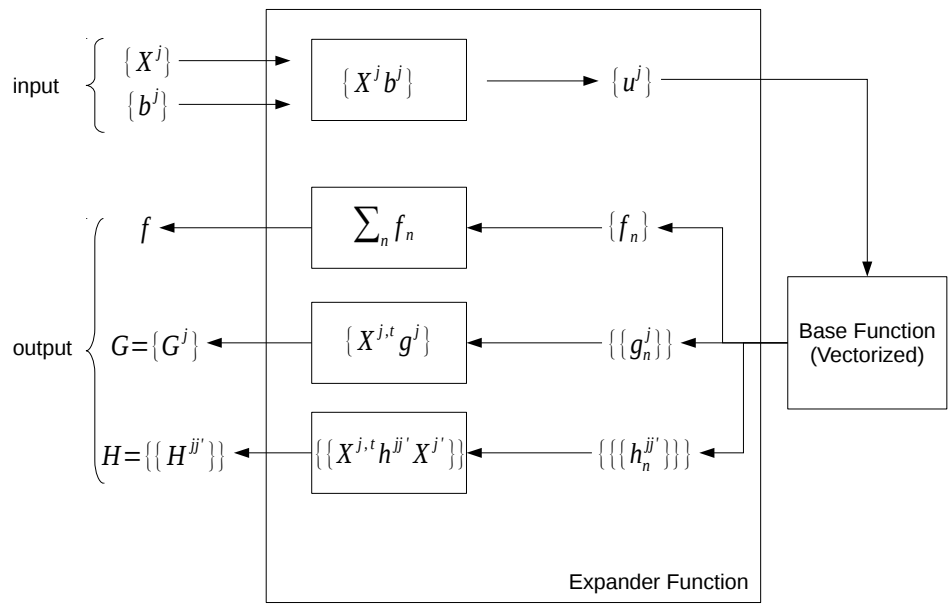}
\caption[]{System flow diagram for \pkg{RegressionFactory}. The expander function is responsible for calculation of log-likelihood and its gradient and Hessian in the high-dimensional space of regression coefficients. It does so by calculating the linear predictors and supplying them to the base function, which is responsible for calculation of log-likelihood and its gradient and Hessian for each data point, in the low-dimensional space of the underlying probability distribution. The expander function converts these low-dimensional objects into the high-dimensional forms, using generic matrix-algebra operations.}.
\label{figure-flow-diagram}
\end{figure}

\pkg{RegressionFactory} is a direct implementation of compact expressions in Equations~\ref{eq-gradient} and \ref{eq-hessian}. These expressions imply a code refactoring by separating model-specific steps (calculation of $\g^j$ and $\hh^{jj'}$) from generic steps (calculation of linear predictors $\XX^j \bbeta^j$ as well as $\XX^{j,t} \g^j$ and $\XX^{j,t} \hh^{jj'} \XX^{j'}$). This decomposition is captured diagramatically in the system flow diagram of Figure~\ref{figure-flow-diagram}.

\subsection{Expander functions}\label{subsection-expanders}
Current implementation of \pkg{RegressionFactory} contains expander and base functions for one-parameter and two-parameter distributions. This covers the majority of interesting GLM cases, and a few more. A notable exception is multinomial regression models (such as logit and probit) which can have an unspecified number of slots. The package can be extended in the future to accommodate such more general cases.

\subsubsection{Single-parameter expander function}\label{subsubsection-expanders-1d}
Below is the source code for \code{regfac.expand.1par}:
\begin{Schunk}
\begin{Sinput}
R> regfac.expand.1par <- function(beta, X, y, fbase1, fgh = 2, ...) {
+    # obtain base distribution derivatives
+    ret <- fbase1(X %*% beta, y, fgh, ...)
+    # expand base derivatives
+    f <- sum(ret$f)
+    if (fgh == 0) return (f)
+    g <- t(X) %*% ret$g
+    if (fgh == 1) return (list(f = f, g = g))
+    xtw <- 0*X
+    for (k in 1:ncol(X)) xtw[, k] <- X[, k] * ret$h
+    h <- t(xtw) %*% X
+    return (list(f = f, g = g, h = h))
+  }
\end{Sinput}
\end{Schunk}
\code{beta} is the vector of coefficients, \code{X} is the matrix of covariates, \code{y} is the vector (or matrix) of response variable, \code{fbase1} is the single-parameter base function being expanded, and \code{fgh} is a flag indicating whether the gradient or Hessian must be returned or not. The \code{dots} argument (\code{...}) is used for passing special, fixed arguments such as the number of trials in a binomial regression. The vectorized function \code{fbase} is expected to return a list of three vectors: \code{f}, \code{g} and \code{h}, corresponding to the base distribution, its first derivative and its second derivative (all vectors of length $N$ or \code{nrow(X)}). The second and third elements correspond to $\g^1$ and $\hh^{11}$ in our notation. Several design aspects of the code are noteworthy for computational efficiency:
\begin{enumerate}
\item Since $\hh$ is diagonal, we only need to return the $N$ diagonal elements.
\item For the same reason, rather than multiplying $\hh$ by $\XX$, we only multiply the vector of diagonal elements by each of the $K$ columns of $\XX$.
\item The flag \code{fgh} controls whether a) only the function value must be returned (\code{fgh==0}), b) only the function and its first derivative must be returned (\code{fgh==1}), or c) the function as well as its first and second derivative must be returned (\code{fgh==2}). This allows optimization or sampling algorithms that do not the first or second derivative to avoid paying an unnecessary computational penalty. Since most often a higher-level derivative implies the need for lower-level derivative(s) (including the function as zero'th derivative), and also since the computational cost of higher derivatives is much higher, the \code{fgh} flag works in an incremental fashion (only 3 options) rather than covering all permutations of \code{f,g,h}.
\end{enumerate}

\subsubsection{Two-parameter expander function}\label{subsubsection-expanders-2d}
Below is the source code for \code{regfac.expand.2par}, the 2D expander function in \pkg{RegressionFactory}:
\begin{Schunk}
\begin{Sinput}
R> regfac.expand.2par <- function(coeff, X
+    , Z=matrix(1.0, nrow = nrow(X), ncol = 1)
+    , y, fbase2, fgh = 2, block.diag = FALSE
+    , ...) {
+    # extracting coefficients of X and Z
+    K1 <- ncol(X); K2 <- ncol(Z)
+    beta <- coeff[1:K1]
+    gamma <- coeff[K1 + 1:K2]
+    
+    # obtain base distribution derivatives
+    ret <- fbase2(X %*% beta, Z %*% gamma, y, fgh, ...)
+  
+    # expand base derivatives
+    # function
+    f <- sum(ret$f)
+    if (fgh == 0) return (f)
+    # gradient
+    g <- c(t(X) %*% ret$g[, 1], t(Z) %*% ret$g[, 2])
+    if (fgh == 1) return (list(f = f, g = g))
+    # Hessian
+    h <- array(0, dim=c(K1+K2, K1+K2))
+    # XX block
+    xtw <- 0 * X
+    for (k in 1:K1) xtw[, k] <- X[, k] * ret$h[, 1]
+    h[1:K1, 1:K1] <- t(xtw) %*% X
+    # ZZ block
+    ztw <- 0 * Z
+    for (k in 1:K2) ztw[, k] <- Z[, k] * ret$h[, 2]
+    h[K1 + 1:K2, K1 + 1:K2] <- t(ztw) %*% Z
+  	# XZ and ZX blocks
+    if (!block.diag) {
+  	  ztw2 <- 0*Z
+  	  for (k in 1:K2) ztw2[,k] <- Z[,k]*ret$h[,3]
+  	  h[K1 + 1:K2, 1:K1] <- t(ztw2)%*%X
+  	  h[1:K1, K1 + 1:K2] <- t(h[K1 + 1:K2, 1:K1])
+  	}
+  	
+    return (list(f = f, g = g, h = h))
+  }
\end{Sinput}
\end{Schunk}
Aside from the same performance optimization techniques used for the one-parameter expander function, the two-parameter expander function has an additional parameter, \code{block.diag}. When \code{TRUE} it sets the cross-derivative terms between the two slots to zero. It can be useful in two scenarios: 1) When the full Hessian is not negative definite, but the Hessian for each parameter is. Block-diagonalization allows for optimization and sampling techniques that rely on this property to be used, at the expense of potentially slower convergence since the block-diagonalized Hessian is not accurate, 2) When optimization of one slot can proceed without knowledge of the value of the other slot, as in many two-parameter exponential family members where the dispersion parameter can be ignored in ML estimation of the mean parameter (e.g. in normal distribution).

\subsection{Base distributions}\label{subsection-base-dist}
Corresponding to the one-parameter and two-parameter expander functions, \pkg{RegressionFactory} offers many of the standard base distributions used in GLM models. Using the nomenclature of \code{glm}, current version (\code{0.7.1}) contains the following base distributions and link functions (* indicates distributions not included in \code{glm} software):
\begin{itemize}
\item One-parameter distributions:
\begin{itemize}
\item Binomial (logit, probit, cauchit, cloglog)
\item Poisson (log)
\item Exponential (log) (*)
\item Geometric (logit) (*)
\end{itemize}
\item Two-parameter distributions:
\begin{itemize}
\item Gaussian (identity / log)
\item Inverse Gaussian (log / log)
\item Gamma (log / log)
\end{itemize}
\end{itemize}

A few points are worth mentioning regarding the choice of base distributions and link functions:
\begin{enumerate}
\item Naming convention: We generally follow this convention for single-parameter distributions:

\code{fbase1.<distribution>.<mean link function>}

and this convention for two-parameter distributions:

\code{fbase2.<distribution>.<mean link function>.<dispersion link function>}

There are can be exceptions. For example, in geometric regression

\code{fbase1.geometric.logit}

the linear predictor is assumed to be \code{logit} of the sucess probability, which is inverse of the distribution mean. Thus, technically the link function is \code{-log(mu-1)}, but for brevity we simply refer to this link function as \code{logit}. Ultimately, naming conventions are less important than the definition of log-likelihood function, which combines the distribution and the link functions.

\item Since the focus of \pkg{RegressionFactory} is on supporting optimization and sampling algorithms for GLM-like models, we are not interested in constant terms in the log-likelihood, i.e. terms that are independent of the regression coefficients. Therefore, we can omit them from the base functions for computational efficiency. An example is the log-factorial term in the Poisson base distribution. (Note that such constant terms are automatically differentiated out of the gradient and Hessian.) If needed, users can implement thin wrappers around the base functions to add the constant terms to the log-likelihood.
\item Our preference is to choose link functions that map the natural domain of the distribution parameter to the real space. For example, in Poisson distribution the natural domain of the distribution mean is the positive real space. The \code{log} link function maps this natural domain to the entire real space. However, for \code{identity} and \code{sqrt} link functions the range is positive real space.
\item We also prefer link functions that produce negative-definiteness for the entire Hessian, or at least for Hessian blocks (corresponding to a subset of the base-distribution parameters). This allow for more optimization/sampling algorithms that take advantage of concavity to be applied to the expanded log-likelihood (according to Theorem~\ref{theorem:concav_1}).
\item We have chosen to absorb the link functions into the function names and their implementation, rather than making distribution names and lonk functions parameters of a single base function. Doing the latter is certainly possible, offering usability at computational cost. Our current choice is driven by the fact that the primary target of \pkg{RegressionFactory} is developers rather than end-users.
\end{enumerate}

\section[]{Using \pkg{RegressionFactory}}\label{section-using}
The most basic application of \pkg{RegressionFactory} is to use the readily-available log-likelihood functions and derivatives. For example, one might be developing a Bayesian model where the log-likelihood is combined with the prior to form the posterior, which is then supplied to a sampling algorithm. Or one might be working on a new optimization algorithm and would like to test its correctness and performance on regression log-likelihood functions as an important use-case. Users can also supply their own base functions to the expander functions of \pkg{RegressionFactory} and readily obtain the log-likelihood and its derivatives. Implementation of functions for calculating base distribution derivatives is often quite simple, which can significantly reduce the time needed for prototyping a new regression model.

There are two equivalent approaches for passing the log-likelihood functions to an optimization/sampling routine: 1) Pass the expander function as the primary function, and the base function as an argument of the primary function, 2) write a thin wrapper that combines the expander and base functions, and pass this wrapper function to the optimization/sampling routine. If the log-likelihood function must be added to another function (such as a prior), then the second approach is the only option where the wrapper implements the logic for adding the two functions. Due to its higher versatility as well as higher code readability, we recommend the second approach.

The above point as well as other usage details are illustrated below, with several examples from single-parameter and double-parameter distributions.

\subsection{Example 1: Bayesian GLM}
The easiest way to take advantage of \pkg{RegressionFactory} is to utilize its standard GLM base functions in custom applications, either for testing the performance of a new optimization/sampling technique, or for composing more complex models from these lego blocks. In the first example, we show how a Bayesian GLM can be constructed in the \pkg{RegressionFactory} framework.

We begin with a basic implementation of Bayesian logistic regression using flat normal priors on each coefficient. First we must load the package into our \proglang{R} session:
\begin{Schunk}
\begin{Sinput}
R> library(RegressionFactory)
\end{Sinput}
\end{Schunk}
Log-likelihood for logistic regression can be readily constructed by applying the single-parameter expander function to the binomial base function and setting the number of trials equal to \code{1}:
\begin{Schunk}
\begin{Sinput}
R> loglike.logistic <- function(beta, X, y, fgh) {
+    regfac.expand.1par(beta, X, y, fbase1.binomial.logit, fgh, n=1)
+  }
\end{Sinput}
\end{Schunk}
We also need a prior for \code{beta}, which we assume to be a normal distribution on each of the \code{K} elements of \code{beta} with the same mean (\code{mu.beta}) and standard deviation (\code{sd.beta}):
\begin{Schunk}
\begin{Sinput}
R> logprior.logistic <- function(beta, mu.beta, sd.beta, fgh) {
+    f <- sum(dnorm(beta, mu.beta, sd.beta, log=TRUE))
+    if (fgh==0) return (f)
+    g <- -(beta-mu.beta)/sd.beta^2
+    if (fgh==1) return (list(f=f, g=g))
+    h <- diag(-1/sd.beta^2, nrow=length(beta))
+    return (list(f=f, g=g, h=h))
+  }
\end{Sinput}
\end{Schunk}
We can now combine the likelihood and prior according to Bayes rule to construct the log-posterior:
\begin{Schunk}
\begin{Sinput}
R> logpost.logistic <- function(beta, X, y, mu.beta, sd.beta, fgh) {
+    ret.loglike <- loglike.logistic(beta, X, y, fgh)
+    ret.logprior <- logprior.logistic(beta, mu.beta, sd.beta, fgh)
+    regfac.merge(ret.loglike, ret.logprior, fgh=fgh)
+  }
\end{Sinput}
\end{Schunk}
In the above, we have taken advantage of the utility function \code{regfac.merge} for combining two lists containing function values and its first two derivatives.

In order to test the above posterior function, we simulate some data using the generative model for logistic regression and estimate the coefficients using \code{glm} for reference:
\begin{Schunk}
\begin{Sinput}
R> N <- 1000
R> K <- 5
R> X <- matrix(runif(N*K, min=-0.5, max=+0.5), ncol=K)
R> beta <- runif(K, min=-0.5, max=+0.5)
R> y <- rbinom(N, size = 1, prob = 1/(1+exp(-X%*%beta)))
R> beta.glm <- glm(y~X-1, family="binomial")$coefficients
\end{Sinput}
\end{Schunk}
We now draw \code{1000} MCMC samples from the posterior of \code{beta} using Stochastic Newton Sampler (SNS), via \proglang{R} package \pkg{sns}~\citep{mahani2014sns}. We are taking advantage of the fact that the sum of two negative-definite Hessians is also negative-definite, a condition needed by SNS. Also, we assume that \code{mu.beta} and \code{sd.beta} are both given to provide a non-informative prior on \code{beta}. Finally, we run \code{sns} in non-stochastic mode via the flag \code{rnd=FALSE} to allow for better comparison of output with \code{glm}:
\begin{Schunk}
\begin{Sinput}
R> library(sns)
R> # for more accurate results and better comparison, increase nsmp
R> nsmp <- 100
R> mu.beta <- 0.0
R> sd.beta <- 1000
R> beta.smp <- array(NA, dim=c(nsmp,K)) 
R> beta.tmp <- rep(0,K)
R> for (n in 1:nsmp) {
+    beta.tmp <- sns(beta.tmp, fghEval=logpost.logistic, X=X, y=y
+      , mu.beta=mu.beta, sd.beta=sd.beta, fgh=2, rnd=FALSE)
+    beta.smp[n,] <- beta.tmp
+  }
R> beta.sns <- colMeans(beta.smp[(nsmp/2+1):nsmp,])
R> cbind(beta.glm, beta.sns)
\end{Sinput}
\begin{Soutput}
      beta.glm    beta.sns
X1  0.01728161  0.01728161
X2 -0.57629725 -0.57629722
X3 -0.55204361 -0.55204359
X4  0.23282848  0.23282846
X5 -0.06941221 -0.06941221
\end{Soutput}
\end{Schunk}
Next, we consider a less-trivial example. We create a hierarchical structure where coefficients of \code{J} groups are assumed to be pooled from normal distribution. This is a simple example of Hierarchical Bayesian models, which due to lack of explanatory variables at the upper level is reduced to a random-coefficient model. We begin with data generation to provide the reader with a tangible grasp of the assumed generative model:
\begin{Schunk}
\begin{Sinput}
R> J <- 20
R> mu.beta.hb <- runif(K, min=-0.5, max=+0.5)
R> sd.beta.hb <- runif(K, min=0.5, max=1.0)
R> X.hb <- list()
R> y.hb <- list()
R> beta.hb <- array(NA, dim=c(J,K))
R> for (k in 1:K) {
+    beta.hb[,k] <- rnorm(J, mu.beta.hb[k], sd.beta.hb[k])
+  }
R> for (j in 1:J) {
+    X.hb[[j]] <- matrix(runif(N*K, min=-0.5, max=+0.5), ncol=K)
+    y.hb[[j]] <- rbinom(N, size=1, prob=1/(1+exp(-X%*%beta.hb[j,])))
+  }
\end{Sinput}
\end{Schunk}
Again, we generate \code{glm} coefficient estimates for reference. Note that \code{glm} treats the groups completely independently of each other, i.e. without any pooling:
\begin{Schunk}
\begin{Sinput}
R> beta.glm.all <- array(NA, dim=c(J,K))
R> for (j in 1:J) {
+    beta.glm.all[j,] <- glm(y.hb[[j]]~X.hb[[j]]-1
+      , family="binomial")$coefficients
+  }
\end{Sinput}
\end{Schunk}
Again, we draw samples from posterior on coefficients using SNS, turning the \code{rnd} flag off for better comparison. Also for code brevity and maintaining focus on how to use pkg{RegressionFactory}, we ignore sampling from the posterior of \code{mu.beta} and \code{sd.beta}, and assume their value is given. We must first construct the log-posteriors. Note that we do not need to change the definition of log-posteior, but the interpretation of \code{mu.beta} and \code{sd.beta} has changed fronm scalars to vector of length \code{K} each:
\begin{Schunk}
\begin{Sinput}
R> beta.smp.hb <- array(NA, dim=c(nsmp,J,K)) 
R> beta.tmp.hb <- array(0.0, dim=c(J,K))
R> for (n in 1:nsmp) {
+    for (j in 1:J) {
+      beta.tmp.hb[j,] <- sns(beta.tmp.hb[j,], fghEval=logpost.logistic
+        , X=X.hb[[j]], y=y.hb[[j]]
+        , mu.beta=mu.beta.hb, sd.beta=sd.beta.hb, fgh=2, rnd=F)
+    }
+    beta.smp.hb[n,,] <- beta.tmp.hb
+  }
R> beta.sns.hb <- apply(beta.smp.hb[(nsmp/2+1):nsmp,,], c(2,3), mean)
\end{Sinput}
\end{Schunk}
We have taken advantage of conditional independence [ref] of the coefficients of each group, given the values of \code{mu.beta} and \code{sd.beta}. We examine the coefficients of the first few groups between \code{glm} and HB methods:
\begin{Schunk}
\begin{Sinput}
R> head(beta.glm.all)
\end{Sinput}
\begin{Soutput}
            [,1]        [,2]        [,3]       [,4]        [,5]
[1,]  0.02469046 -0.02179051  0.09795783 -0.3686893  0.37422102
[2,]  0.23716478 -0.14741078 -0.64943037 -0.1177934  0.17251167
[3,] -0.08602828  0.07878448  0.16795872  0.1100804  0.03127816
[4,]  0.00178465  0.24283261 -0.24240170 -0.2504084 -0.15572659
[5,]  0.17083027  0.32218840  0.13358916  0.1168579 -0.15588975
[6,]  0.22690633  0.11722254 -0.08157556  0.3395405  0.02823816
\end{Soutput}
\begin{Sinput}
R> head(beta.sns.hb)
\end{Sinput}
\begin{Soutput}
             [,1]        [,2]        [,3]        [,4]        [,5]
[1,] -0.008028756 -0.03689594  0.13273096 -0.36892225  0.27940600
[2,]  0.166836113 -0.15122282 -0.54964963 -0.14652433  0.10439417
[3,] -0.103947649  0.05881464  0.19355822  0.05347266 -0.02005739
[4,] -0.029730149  0.21519779 -0.17819428 -0.26395210 -0.18503945
[5,]  0.122504007  0.29084126  0.15998503  0.06725949 -0.18384662
[6,]  0.169507903  0.10188925 -0.03573334  0.26120405 -0.02598995
\end{Soutput}
\end{Schunk}
Plotting unpooled (\code{glm}) and pooled (\code{sns}) coefficients shows the typical shrinkage pattern of Bayesian models.
\begin{Schunk}
\begin{Sinput}
R> plot(beta.glm.all[,1], beta.sns.hb[,1]
+       , xlab="Unpooled Coefficients"
+       , ylab="Pooled Coefficients")
R> abline(a=0, b=1)
\end{Sinput}
\end{Schunk}

\begin{figure}
\includegraphics{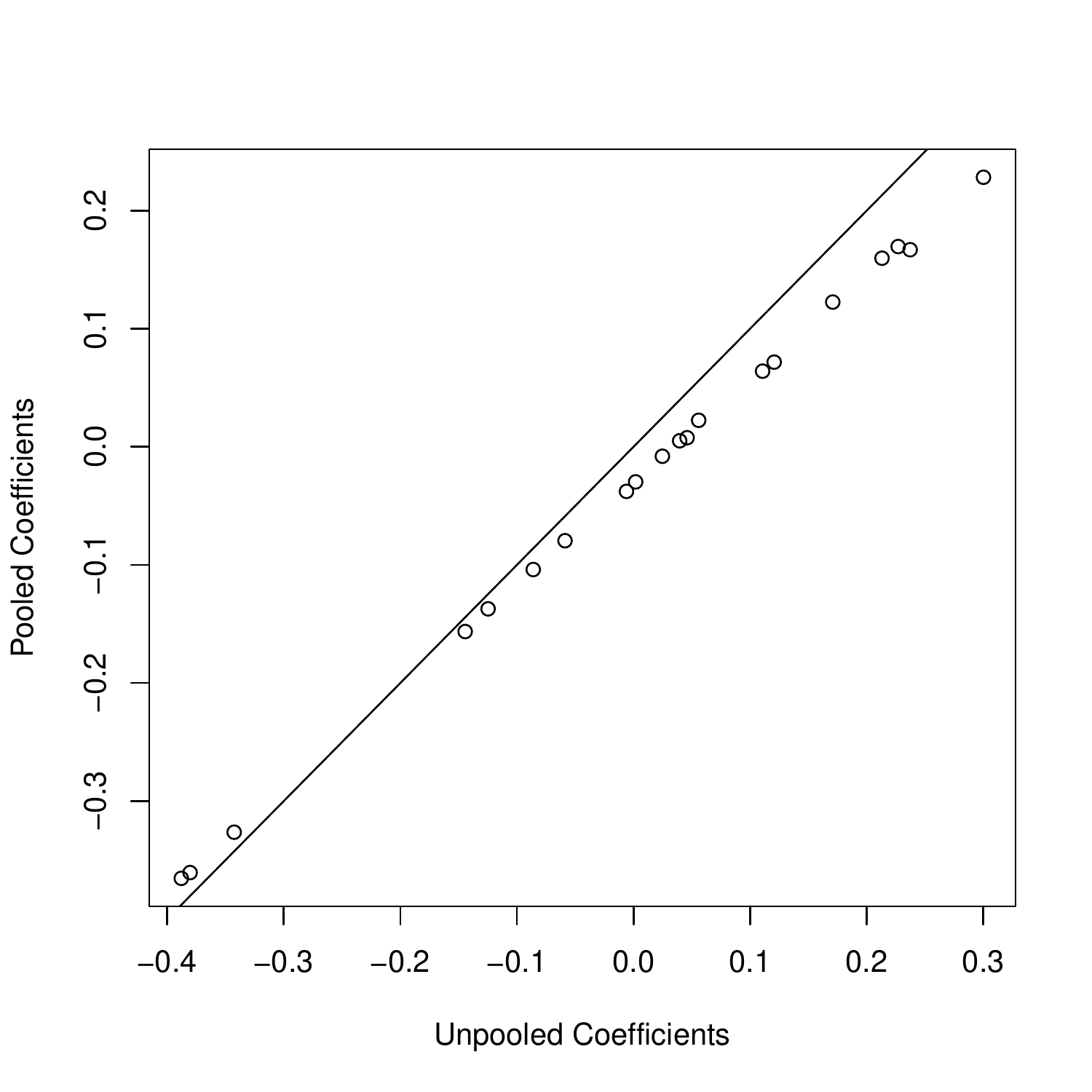}
\caption{Pooling of logistic regression coefficients using a hierarchical Bayesian framework produces the familiar shrinkage towards the mean effect.}
\label{fig-shrinkage}
\end{figure}

\subsection{Example 2: Double-parameter GLM with varying dispersion}
As a second example, we consider a double-parameter GLM with varying dispersion, i.e., dependent on the covariates. As of version \code{0.7.1}, \pkg{RegressionFactory} contains three double-parameter base distributions: Gaussian, inverse Gaussian, and Gamma. These double-parameter distributions can be used in a constant-dispersion or varying-dispersion setting. Constant-dispersion scenario is a special case of varying-dispersion scenario where the only covariate used to explain the dispersion parameter of the base distribution is intercept. This corresponds to the default value of \code{Z} in the function \code{regfac.expand.2par}.

First, we load the \proglang{R} package \pkg{dglm}:
\begin{Schunk}
\begin{Sinput}
R> library(dglm)
\end{Sinput}
\end{Schunk}
To use \pkg{RegressionFactory}, as before we implement a thin wrapper to combine the 2D expander with the normal base distribution:
\begin{Schunk}
\begin{Sinput}
R> loglike.linreg <- function(coeff, X, y, fgh, vd = F) {
+    if (vd) regfac.expand.2par(coeff = coeff, X = X, Z = X, y = y
+      , fbase2 = fbase2.gaussian.identity.log, fgh = fgh, block.diag = F)
+    else regfac.expand.2par(coeff = coeff, X = X, y = y
+      , fbase2 = fbase2.gaussian.identity.log, fgh = fgh, block.diag = F)
+  }
\end{Sinput}
\end{Schunk}
The boolean flag \code{vd} indicates whether we want to use covariates to explain the dispersion or not. If \code{FALSE}, the model is reduced to ordinary linear regression. Next, we simulate data according to the assumed generative model:
\begin{Schunk}
\begin{Sinput}
R> N <- 1000
R> K <- 5
R> X <- matrix(runif(N*K, min=-0.5, max=+0.5), ncol=K)
R> beta <- runif(K, min=-0.5, max=+0.5)
R> gamma <- runif(K, min=-0.5, max=+0.5)
R> mean.vec <- X%*%beta
R> sd.vec <- exp(X%*%gamma)
R> y <- rnorm(N, mean.vec, sd.vec)
\end{Sinput}
\end{Schunk}
We now estimate constant-dispersion and varying-dispersion models using the \proglang{R} commands \code{lm} and \code{dglm}:
\begin{Schunk}
\begin{Sinput}
R> # constant-dispersion model
R> est.glm <- lm(y~X-1)
R> beta.glm <- est.glm$coefficients
R> sigma.glm <- summary(est.glm)$sigma
R> # varying-dispersion model
R> est.dglm <- dglm(y~X-1, dformula = ~X-1, family = "gaussian", dlink = "log")
\end{Sinput}
\begin{Soutput}
family: gaussian 
\end{Soutput}
\begin{Sinput}
R> beta.dglm <- est.dglm$coefficients
R> gamma.dglm <- est.dglm$dispersion.fit$coefficients
\end{Sinput}
\end{Schunk}
Finally, we estimate the same models using the expander framework of \pkg{RegressionFactory}:
\begin{Schunk}
\begin{Sinput}
R> # constant-dispersion
R> coeff.smp <- array(NA, dim=c(nsmp, K+1)) 
R> coeff.tmp <- rep(0, K+1)
R> for (n in 1:nsmp) {
+    coeff.tmp <- sns(coeff.tmp, fghEval=loglike.linreg
+      , X=X, y=y, fgh=2, vd = F, rnd = F)
+    coeff.smp[n,] <- coeff.tmp
+  }
R> beta.sns.cd <- colMeans(coeff.smp[(nsmp/2+1):nsmp, 1:K])
R> sigma.sns.cd <- sqrt(exp(mean(coeff.smp[(nsmp/2+1):nsmp, K+1])))
R> cbind(beta.glm, beta.sns.cd)
\end{Sinput}
\begin{Soutput}
     beta.glm beta.sns.cd
X1 -0.2630246  -0.2630246
X2 -0.4988430  -0.4988430
X3  0.5966928   0.5966928
X4 -0.3255354  -0.3255354
X5  0.1037526   0.1037526
\end{Soutput}
\begin{Sinput}
R> cbind(sigma.glm, sigma.sns.cd)
\end{Sinput}
\begin{Soutput}
     sigma.glm sigma.sns.cd
[1,]  1.031139     1.028558
\end{Soutput}
\begin{Sinput}
R> # varying-dispersion
R> coeff.smp <- array(NA, dim=c(nsmp, 2*K)) 
R> coeff.tmp <- rep(0, 2*K)
R> for (n in 1:nsmp) {
+    coeff.tmp <- sns(coeff.tmp, fghEval=loglike.linreg
+      , X=X, y=y, fgh=2, vd = T, rnd = F)
+    coeff.smp[n,] <- coeff.tmp
+  }
R> beta.sns.vd <- colMeans(coeff.smp[(nsmp/2+1):nsmp, 1:K])
R> gamma.sns.vd <- colMeans(coeff.smp[(nsmp/2+1):nsmp, K+1:K])
R> cbind(beta.dglm, beta.sns.vd)
\end{Sinput}
\begin{Soutput}
    beta.dglm beta.sns.vd
X1 -0.2702744  -0.2702757
X2 -0.5500612  -0.5500678
X3  0.6352290   0.6352354
X4 -0.3866443  -0.3866514
X5  0.1004077   0.1004075
\end{Soutput}
\begin{Sinput}
R> cbind(gamma.dglm, gamma.sns.vd)
\end{Sinput}
\begin{Soutput}
   gamma.dglm gamma.sns.vd
X1  0.3183887    0.3184326
X2  0.5214325    0.5214858
X3  0.2689971    0.2689735
X4  0.6318339    0.6318852
X5 -0.4145652   -0.4145799
\end{Soutput}
\end{Schunk}
Note that the mean coefficients from \pkg{dglm} and \pkg{RegressionFactory} match exactly in constant-dispersion case, but the dispersion parameters do not match since \code{dglm} uses a method of moments to estimate dispersion, rather than log-likelihood maximization. For varying-dispersion scenario, since mean and dispersion coefficients are estimated simultaneously, neither sets match exactly between the two methods, but they are very close, and the discrepancy becomes smaller for larger data.

\subsection{Example 3: Geometric regression}\label{subsection-geometric}
In the last example, we illustrate how a new GLM regression can be easily constructed using the \pkg{RegressionFactory} framework. This involves three steps: 1) identify a base distribution, 2) select the link function(s), and 3) combine 1 and 2 to arrive at the log-likelihood function and its derivatives, preferrably to make the Hessian negative-definite. According to Theorem~\ref{theorem:concav_1}, this property can be proven in the base-distribution space, which is often quite easy. Consider the geometric distribution:
\begin{equation}
P(y=k; p) = (1-p)^{k-1}p.
\end{equation}
Using a logit link function for $p$, we arrive at the following log-likelihood:
\begin{equation}
f(u; y) = - \left( y \, u + (1 + y) \, \log {1 + e^{-u}} \right).
\end{equation}
Concavity of the above function can be easily verified:
\begin{equation}
f_{uu} = -(1 + y) e^u / (1 + e^u)^2 < 0
\end{equation}
The base function \code{fbase1.goemetric.logit} implements the above log-likelihood and its first two derivatives. To test the function, we first simulate data from the distribution:
\begin{Schunk}
\begin{Sinput}
R> N <- 1000
R> K <- 5
R> X <- matrix(runif(N*K, min=-0.5, max=+0.5), ncol=K)
R> beta <- runif(K, min=-0.5, max=+0.5)
R> y <- rgeom(N, prob = 1/(1+exp(-X%*%beta)))
\end{Sinput}
\end{Schunk}
We now use SNS in non-stochastic mode (i.e. Newton optimization) to estimate the coefficients. We begin by our usual thin wrapper around the expander function to fully implement the log-likelihood.
\begin{Schunk}
\begin{Sinput}
R> loglike.geometric <- function(beta, X, y, fgh) {
+    regfac.expand.1par(beta, X, y, fbase1.geometric.logit, fgh)
+  }
R> beta.est <- rep(0,K)
R> for (n in 1:10) {
+    beta.est <- sns(beta.est, fghEval=loglike.geometric
+      , X=X, y=y, fgh=2, rnd = F)
+  }
R> cbind(beta, beta.est)
\end{Sinput}
\begin{Soutput}
           beta   beta.est
[1,]  0.3631200  0.4260850
[2,]  0.1219817  0.1695647
[3,] -0.3779823 -0.4790984
[4,] -0.4623841 -0.2790712
[5,] -0.1732932 -0.3219395
\end{Soutput}
\end{Schunk}

\section{Summary}\label{section-summary}
We presented \proglang{R} package \pkg{RegressionFactory}, a modular framework for evaluating GLM log-likelihood functions and their derivatives. We illustrated its utility in rapidly developing composite GLM models such as Hierarchical Bayesian as well as new regression models such as geometric and exponential regression. The accompanying definiteness-invariance theorem allows us to reason about logl-likelihood Hessian in a much lower-dimensional space.

Another advantage of our modular implementation is that it allows for performance optimization strategies to be readily applied across all GLM models. For example, the linear algebra steps contained in the expansion functions \code{regfac.expand.1par} and \code{regfac.expand.2par} can be thoroughly studied from the following perspectives:
\begin{itemize}
\item Row-major vs. column-major layout of the covariate matrices $\XX^j$'s, for single-threaded and multi-threaded scenarios.
\item Non-Uniform Memory Access (NUMA) implications of memory allocation for $\XX^j$'s.
\item Loop and cache fusion strategies.
\item Coarse- vs. fine-grained parallelization in composite models such as HB.
\end{itemize}
While base functions contain model-specific code, yet they also present broad optimization opportunities. For example, they are all vectorized by definition, suggesting that they can benefit from optimized Single-Instruction, Multiple-Data (SIMD) implementation. In particular, access to vectorized transcendental functions can greatly improve the peformance of many base functions. Many of the above issues have been studied here~\citep{mahani2013simd}. A natural next step for \pkg{RegressionFactory} would be to implement the expander and base functions in compiled code such as \proglang{C/C++}, which allows for many of the advanced optimization techniques to be applicable subsequently.


\bibliography{RegressionFactory}
\end{document}